\documentclass{elsart}

\usepackage{graphicx}
\usepackage{amssymb}

\begin{document}

\begin{frontmatter}

\title{A note on  non-thermodynamical applications of
non-extensive statistics}

\author{Dami\'an H. Zanette}
\ead{zanette@cab.cnea.gov.ar}

\address{Consejo Nacional de Investigaciones Cient\'{\i}ficas y
T\'ecnicas, Centro At\'omico Bariloche and Instituto Balseiro,
8400 Bariloche, R\'{\i}o Negro, Argentina}

\author{Marcelo A. Montemurro}
\ead{mmontemu@ictp.trieste.it}

\address{Abdus Salam International Centre for Theoretical Physics,
Strada Costiera 11, 34100 Trieste, Italy}

\begin{abstract}
It is pointed out that the constraint to be imposed to the
maximization of the entropy for processes outside the class of
thermodynamical systems, is generally not well defined. In fact,
{\it any} probability distribution can be derived from Jaynes's
principle with a suitable choice of the constraint. In the case
of Tsallis's non-extensive formalism, this implies that it is not
possible to establish any connection between specific
non-thermodynamical processes and non-extensive mechanisms and,
in particular, to assign any unambiguous non-extensivity index
$q$ to those processes.
\end{abstract}


\end{frontmatter}

Boltzmann-Gibbs thermostatistics bridges the microscopic
description of physical systems that obey the laws of mechanics
with the macroscopic picture drawn from the principles of
thermodynamics. From a historical perspective, it reconciled the
physics of thermal processes, fully congenial with our
everyday-life experience, and the mechanistic interpretation of
the Universe as a huge ensemble of interacting particles --two
views that, in the middle of the nineteenth century, were far
from being perceived as compatible, both mathematically and
philosophically \cite{Cerci}.

Physical systems whose microscopic degrees of freedom are
governed by the equations of (classical or quantum) mechanics,
and which have reached a state of macroscopic equilibrium, will
hereafter be called {\it thermodynamical systems}. The
Boltzmann-Gibbs formulation makes it possible to derive a theory
for the collective equilibrium state variables of a
thermodynamical system from its microscopic dynamics. At the same
time, it gives origin to the mesoscopic (statistical) description
level. An overwhelming corpus of experimental work validates the
results of this procedure, both at mesoscopic and macroscopic
levels. It is known, however, that Boltzmann-Gibbs
thermostatistics fails to give a mathematically consistent
description of certain physical systems --notably, those driven by
gravitational and other long-range interactions-- since the
predicted values of some of their state variables diverge
\cite{saslaw}. Though it is not clear whether such systems attain
at all a state identifiable with macroscopic equilibrium
\cite{tremaine}, their thermodynamical properties would be
characterized by non-extensive macroscopic variables.

In 1988, C.~Tsallis proposed an extension of Boltzmann-Gibbs
equilibrium thermostatistics based on a variational principle for a
generalized form of the entropy \cite{TS1},
\begin{equation} \label{S}
S_q=- \frac{\sum_i p_i^q-1}{q-1},
\end{equation}
where $p_i$ is the probability that the thermodynamical system
under study is found in its $i$-th quantum state and $q\in
(-\infty,\infty)$. The fact that, for $q\ne 1$, the entropy $S_q$
is not additive with respect to the factorization of the
probabilities, has led to the assumption that this formalism may
provide consistent thermostatistical description of non-extensive
systems. The parameter $q$ has been called non-extensivity index,
since it measures the deviation from additivity of $S_q$.

In the canonical scenario, the generalized entropy (\ref{S}) is
maximized with respect to $p_i$, with the constraint of probability
normalization,
\begin{equation} \label{1}
\sum_i p_i=1,
\end{equation}
and fixing the value of a generalized form of the mean energy
\cite{TS15},
\begin{equation} \label{E}
\frac{\sum_i p_i^q \epsilon_i}{\sum_i p_i^q}=E_q,
\end{equation}
where $\epsilon_i$ is the energy of the $i$-th state. This canonical
maximization procedure yields the energy-dependent probability
distribution
\begin{equation} \label{pi}
p_i=Z^{-1} [1+\beta (q-1) \epsilon_i]^{-1/(q-1)},
\end{equation}
where $Z$ is a normalization constant analogous to the partition
function and $\beta$ is an auxiliary parameter. The variational
formulation of canonical Boltzmann-Gibbs thermostatistics is
fully recovered in the limit $q \to 1$, where $\beta$ reduces to
the inverse temperature.

A remarkable property of Tsallis's generalization is that it
preserves the mathematical structure of standard thermostatistics
for any value of $q$ \cite{TS2}.  This noticeable feature
justifies the rather unexpected form of the constraint (\ref{E}),
which replaces the usual definition of the mean energy $E =\sum_i
p_i\epsilon_i$. In a long series of publications \cite{cbpf}, it
has been shown that most of the theorems of equilibrium
statistical mechanics, as well as many results concerning linear
and nonlinear non-equilibrium properties, can be formally
generalized in the frame of the extended formalism. Apart from
this formal equivalence with Boltzmann-Gibbs thermostatistics,
the relevance of the non-extensive formulation should be
validated by the observation of actual thermodynamical systems
with an energy distribution of the form (\ref{pi}). At the same
time, the role of the quantity $E_q$ as a macroscopic property of
the system in question should be assessed, in comparison with the
role of the mean energy $E$ as a state variable of extensive
systems. Until now, however, there is no conclusive evidence that
non-extensive thermostatistics might correctly describe any
thermodynamical system \cite{ZM1,ZM2,9}.

On the other hand, many real systems have been identified where the
statistical distribution of their relevant variables --not of the
energy, however-- are well fitted with functions of the type of
Eq.~(\ref{pi}) \cite{cbpf}. To be specific, empirical distributions
for quantities $x$ defined over the semi-infinite range $(0,\infty)$
have been systematically fitted with the two-parameter function
\begin{equation} \label{p1}
p(x)= N (1+a x)^b,
\end{equation} [cf. Eq. (\ref{pi})]
where $N(a,b)$ is chosen in such a way that $p(x)$ is normalized
to unity. Through identification of $p(x)$ with the distribution
of Eq. (\ref{pi}), the fitting parameters $a$ and $b$ are used to
assign an ``inverse temperature'' $\beta$ and a non-extensivity
index $q$ to the empirical distribution under consideration. For
quantities $x\in (-\infty,\infty)$ the chosen function is,
instead,
\begin{equation} \label{p2}
p(x)= N (1+a x^2)^b.
\end{equation}
This approach has been applied, for instance, to momentum
distributions in elementary particle interactions \cite{ee},
velocity distributions in diffusing biological systems
\cite{hydra}, and volume and return distributions in financial
processes \cite{econ}. In all the reported cases the result of
the fitting seems to be quite good, a circumstance that has
invariably led to the claim that the systems under study are
governed by mechanisms characterized by non-extensivity.
Frequently, moreover, connections have been established with
presumably related concepts, such as self-similarity, scale
invariance, non-ergodicity, meta- and quasi-equilibria,
criticality, algorithmic complexity, {\it et c{\ae}tera}
\cite{cbpf,econ,etc}.

Now, the fact that an empirical distribution is well fitted by a
function derived from a variational principle analogous to that of
canonical thermostatistics, does not necessarily mean that the
nature of the underlying processes is the same as in a
thermodynamical system, as described by equilibrium statistical
mechanics, nor that the state of the system can be identified
with canonical thermal equilibrium. The existence of a mechanical
Hamiltonian formulation for processes such as, say, elementary
interactions may be a matter of controversy, but a
Hamiltonian-like realistic description of a biological population
or the stock market should be out of question. However obvious,
this remark raises a significant question associated with the
maximization procedure that yields fitting functions such as
those of Eqs. (\ref{p1}) and (\ref{p2}), and with Jaynes's
principle in general: Besides probability normalization, which is
the ``correct'' constraint to be used in the canonical
maximization of the entropy?

For extensive thermodynamical systems, numberless instances of
experimental validation show that the mean energy $E$ is to be
fixed. The formal equivalence of non-extensive themostatistics
with the Boltzmann-Gibbs formulation suggests in turn that
constraint (\ref{E}) may be necessary to deal with non-extensive
thermodynamical systems (see, however, Ref. \cite{ZM2}). On the
other hand,  no rigorous justification can generally hold for any
of the infinitely many constraints that can be imposed in the case
of non-thermodynamical systems \cite{Jaynes}. Assuming that the
states of the system are well defined, any function $\phi(x)$ of
the relevant variable $x$ may be used to introduce the average
\begin{equation} \label{Fi}
\Phi_q = \frac{\sum_i p_i^q \phi(x_i)}{\sum_i p_i^q},
\end{equation}
in full analogy with Eq. (\ref{E}). Maximization of the entropy
(\ref{S}) under constraints (\ref{1}) and (\ref{Fi}) leads to
\begin{equation} \label{pfi}
p(x)= Z^{-1}[1+\beta (q-1) \phi(x)]^{-1/(q-1)}.
\end{equation}
This freedom in the choice of the variational constraint seems to
have been systematically overlooked by those authors who applied
non-extensive thermostatistics to the description of empirical
data from non-thermodynamical systems. In fact, choosing fitting
functions as in Eqs. (\ref{p1}) and (\ref{p2}) amounts to
restricting $\phi(x)$ to $x$ and $x^2$, respectively.

An exception to this rule, which dramatically points out the
necessity of considering more general constraints when studying
non-thermodynamical systems, is given by the fittings of
velocity-difference distributions in turbulent flows
\cite{turbul1,turbul2,turbul3}. In this case, reasonable fittings
with functions as in Eq.~(\ref{pfi}) are obtained only when
$\phi(x)$ is allowed to take rather complicated forms, typically,
$\phi(x)=x^{2\alpha}/2-c\ \mbox{sgn}(x)
(|x|^\alpha-|x|^{3\alpha}/3) $ with $c\ge 0$ and $0<\alpha< 1$.
The circumstance that such form of $p(x)$ implies an unusual
choice of the function being averaged in the variational
constraint does not seem to have been discussed in the relevant
literature, though.

\begin{figure}
\centering \resizebox{\columnwidth}{!}{\includegraphics{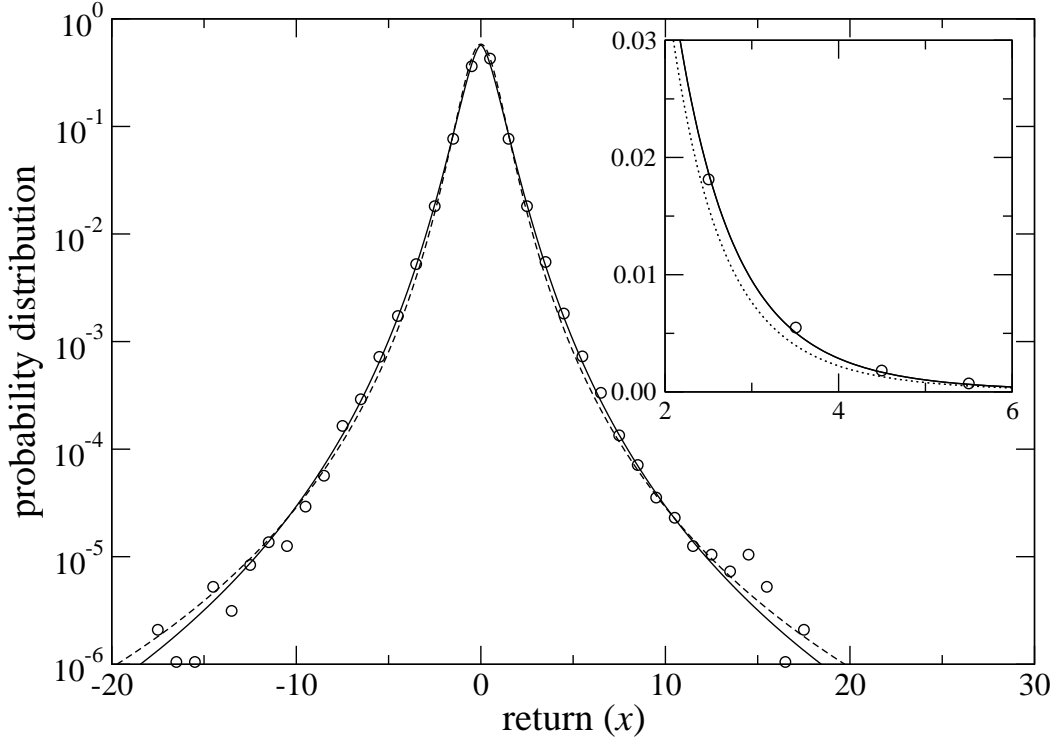}}
\caption{Distribution of (normalized) one-minute returns of $10$
stocks of the New York Stock Exchange during 2001. The dotted line is
a fitting with Eq. (\ref{p2}), which yields $q=1.4$ \cite{econ}. The
full line corresponds to a fitting with Eq. (\ref{pfi}), taking
$\phi(x)= |x|^{1.6}$. In this case, $q=1.3$. The inset shows a
close-up in linear scales.}
\label{fig}
\end{figure}

The arbitrariness of the variational constraint for
non-thermodynamical systems makes it possible to finely tune the
fitting function by a suitable choice of $\phi (x)$. As an
illustration, we show in Fig. \ref{fig} an empirical distribution
of high-frequency stock returns in the New York Stock Exchange
\cite{econ2}, along with two fittings. The dashed curve
corresponds to the fitting function of Eq. (\ref{p2}), with
$b=-2.5$ ($q=1.4$) \cite{econ}. While the overall quality of the
approximation is good, a systematic deviation from the empirical
data is apparent for intermediate values of the distribution
($10^{-4}\lesssim p(x) \lesssim 10^{-2}$). This deviation is
considerably reduced if, as shown by the full curve, the
maximization of entropy is subject to constraint (\ref{Fi}) with
$\phi(x)=|x|^{1.6}$, which gives $q=1.3$. As discussed above,
this constraint is as valid as any other, and has the advantage
of yielding a better fitting for the empirical data.

It immediately results from Eq. (\ref{pfi})
that introducing the function
\begin{equation} \label{fi}
\phi(x) =\frac{1}{q-1}[A p(x)^{1-q}+\phi_0]
\end{equation}
in constraint (\ref{Fi}) leads to a variational principle which
{\it exactly} yields {\it any} given distribution $p(x)$. The
constants $A$ and $\phi_0$ fix the origin and units of measure
for the average $\Phi_q$, while $q$ establishes the connection
between the shapes of $\phi(x)$ and $p(x)$. Note that the index
$q$ can be chosen arbitrarily, and that a different form of
$\phi(x)$ is obtained for each value of $q$. For $q\to 1$, we get
$\phi(x) = A'\ln p(x)+\phi_0'$.

The simple observation that {\it any} distribution can be derived
from a variational principle for the entropy if a suitable
constraint is chosen, has far-reaching consequences when
interpreting the fitting of non-thermodynamical empirical data in
the frame of non-extensive statistics. In particular, it voids of
meaning any claim of connection between the fitted data and
possible non-extensive mechanisms underlying the system in
question. A quantitative proof of this assertion is provided by
the fact that, for a given system, the non-extensivity index $q$
is not uniquely defined, and can in fact be given any value by an
appropriate choice of $\phi(x)$. Any system, in fact, could be
made ``extensive'' by simply using the constraint that yields
$q=1$!

At this point, it could be argued that the kind of constraints
arising from a choice of $\phi(x)$ as in Eq. (\ref{fi}) will
generally be unconventional, typically involving complicated
functions of the relevant variables. We have seen that this is in
fact the case if the distributions of velocity-differences in
turbulent flows are forced to fit non-extensive thermostatistics.
For other physical systems, it has already been remarked that
insisting to stick to a variational principle may necessarily
lead to consider ``non-traditional'' constraints \cite{MS}. For
non-physical systems, unfortunately, we can hardly discern
between ``traditional'' and ``non-traditional'' constraints.
Without a rigorous argument to decide on this point, no choice
can possibly be dismissed on such basis.

In summary, we have pointed out that any probability distribution
can be derived from the entropic variational formalism that
underlies Tsallis's non-extensive thermostatistics, if a suitable
``state function'' is chosen to define a constraint for the
maximization procedure. The same remark should hold for any
variational formalism based on Jaynes's principle. While for
macroscopic Hamiltonian systems in thermodynamical equilibrium it
is well established that the mean energy is to be fixed, for
non-thermodynamical systems it is generally not possible to argue
for or against any choice. Therefore, any claim of connection
between non-thermodynamical processes and non-extensive
mechanisms, based on the fitting of empirical probability
distributions with the functions derived from Tsallis's
variational formalism, is essentially insubstantial. Such kind of
fittings provide at most a phenomenological description of the
systems in question, and bear little information on their true
nature.

\end{document}